# Phase co-existence and structured diffuse scattering seen by X-ray 3D mapping of reciprocal space for $Cs_{0.8}Fe_{1.6}Se_2$.

A. Bosak[2], V. Svitlyk[1], A. Krzton-Maziopa[3], E. Pomjakushina[3], K. Conder[3], V. Pomjakushin[4], A. Popov[2], D. de Sanctis[2], and D. Chernyshov[1]

[1] *Swiss-Norwegian Beamlines at European Synchrotron Radiation Facility, BP 220, 38043 Grenoble, France*

[2] *European Synchrotron Radiation Facility, BP 220, 38043 Grenoble Cedex, France*

[3] *Laboratory for Developments and Methods, PSI, CH-5232 Villigen PSI, Switzerland*

[4] *Laboratory for Neutron Scattering, Paul Scherrer Institut, CH-5232 Villigen PSI, Switzerland*

ABSTRACT

$A_xFe_{2-y}Se_2$ (A=K, Rb, Cs) superconductors may show a complex mixture of structural phases even in a single crystal form. A full sphere of the diffraction intensity has been collected, with the help of synchrotron radiation, for $Cs_{0.8}Fe_{1.6}Se_2$. In addition to the expected pattern for the tetragonal phase with ordered Fe vacancies, a diffuse scattering from Cs occupational disorder has been observed, together with an extra Bragg contribution from a minor phase. The minor phase, in agreement with previous findings, is compressed in the tetragonal *a-b* plane and expanded along the *c*-direction; a set of modulated Bragg rods evidences a planar disorder. Fourfold splitting of the rods as well as the main Bragg peaks for L≠0 imply that symmetry of the minor phase is not higher than *monoclinic*. The monoclinic distortion was estimated to be 90.7 degrees. Structured diffuse scattering, observed on top of the Bragg component, relates to the major phase and is attributed to a correlated distribution of Cs ions. Elastic nature of the observed scattering was confirmed by additional inelastic X-ray scattering experiment. Diffuse scattering forms 3D objects that have topological similarity with what one would expect, under assumptions discussed in the text, for a Fermi surface nesting.

INTRODUCTION

Single crystal diffraction gives the most direct and precise information on the crystal structure. A complexity of the diffraction pattern should therefore correspond to complexity of real structure that, for disordered solids, necessitates a complete collection of the diffracted intensities covering few orders of magnitude from very strong Bragg reflections to very weak diffuse scattering. An advent of synchrotron radiation and modern area detectors with high dynamic range makes these scattering experiments feasible. Here we apply such diffraction technique to study crystal structure of novel $Cs_{0.8}Fe_{1.6}Se_2$ superconductor [1].

Recently a complex multiphase composition has been proposed for $A_xFe_{2-y}Se_2$ family, here A stays for alkaline metal and x and y are close to 0.8 and 0.4 respectively; corresponding diffraction experiments with powder or single crystal samples seem to confirm the multiphase state. Thus, there is general agreement that the major phase is of I/m symmetry, shows (partially) ordered pattern of Fe vacancies and stays magnetically ordered up to above 500K [2, 3]. At high temperatures this major phase transforms to I/mmm disordered phase with no magnetic order [2]. The high temperature structure is often considered as an average one in a sense of average occupancy of the iron positions and is the same as in the layered (122-type) iron pnictides [4]. Symmetry relations between phases have been listed in Ref. [5], here we remind that the low temperature structure has the unit cell $\sqrt{5} \times \sqrt{5}$ bigger than that for the high temperature average version and manifests in superstructure reflections with the propagation vector [2/5 1/5 1].

The presence of a second phase has been noted in many diffraction studies, often proofed by additional reflections with propagation vector [1/2,1/2,0], that is refereed as a "$\sqrt{2} \times \sqrt{2}$ superstructure" of yet unknown composition and crystal structure; the other phases not having such modulation have also been proposed. For example, in the transmission electron microscopy experiment [6], the superstructure $\sqrt{5} \times \sqrt{5}$ was observed together with $\sqrt{2} \times \sqrt{2}$ pattern in certain areas for non-superconducting $K_yFe_{2-x}Se_2$ ($Fe_{1.5-1.6}$, here and below  stays for the vacancy), whereas the superconducting (SC) crystals ($Fe_{1.6-1.7}$) showed a phase-separated state comprising the $\sqrt{5} \times \sqrt{5}$ superstructure and a disordered 122-structure having slightly different *c* lattice dimensions. Scanning tunnelling microscopy study [7] also shows charge density modulation with $\sqrt{2} \times \sqrt{2}$ periodicity, but no $\sqrt{5} \times \sqrt{5}$ superstructure in presumably SC $K_yFe_{2-x}Se_2$, thus proposing a microscopic coexistence of superconductivity and an AFM state without an ordering of Fe-vacancies. In the single crystal synchrotron microfocusing diffraction experiments [8, 9] a phase-separated inhomogeneous state has been observed for superconducting $K_yFe_{2-x}Se_2$ with a majority of $\sqrt{5}\times\sqrt{5}$ phase and a minority (~30%) of a phase having in-plane compressed lattice; $\sqrt{2}\times\sqrt{2}$ weak superstructure has also been reported. Bragg reflections of the in-plane compressed minor phase disappear on heating above 520 K. The compressed phase with $\sqrt{2}\times\sqrt{2}$ superstructure has also been observed in the single crystal neutron diffraction experiments [10]. This phase was tentatively described in Pnma space group with an ordered vacancy model [10]; however a quasi-two-dimensional rod-like character of $\sqrt{2}\times\sqrt{2}$ superstructure reflections has been noted. In our previous work [5] we have also reported the observation of (1/2,1/2,L) diffraction rods together with $\sqrt{5}\times\sqrt{5}$ superstructure in $Cs_yFe_{2-x}Se_2$. In the present paper we apply X-ray 3D mapping of reciprocal space to characterize the minor phase in more details taking $Cs_{0.8}Fe_{1.6}Se_2$ as an example.

The majority of possible structural phases proposed for $A_xFe_{2-y}Se_2$ compounds incorporate a certain amount of disorder related to *x* or *y* stoichiometry. As a result, one could expect a diffuse scattering related predominantly to an occupational disorder in A and Fe sublattices. This not yet probed scattering component would carry information on the deviations from average structure caused by fluctuations of composition and corresponding lattice distortions. Fluctuation spectrum relates to an effective interaction potential and diffuse scattering therefore may also be used to map this potential in the reciprocal space [11]. Diffuse scattering related to the perturbation of the potential by nesting effect [12, 13, 14, 15] may be linked to the topology of the Fermi surface; similar to the problem of phase coexistence, the necessary experimental information can be obtained with 3D mapping of reciprocal space. We present such maps for $Cs_{0.8}Fe_{1.6}Se_2$ that show structured diffuse scattering, mainly associated with a disorder in Cs-sublattice, and discuss topology of the observed diffuse objects.

First part of the paper presents necessary experimental information related to the sample, the data collection technique and post-experimental data treatment. Second one contains discussion of Bragg scattering component and assignment of the observed diffraction features to various phases on the basis of translational commensurability. In third part, we consider diffuse scattering component and parameterize it with a set of short-range correlation parameters. Forth part comprises detailed description of the observed diffuse objects topology with comments on its similarity to a nesting construction.

EXPERIMENTAL

Single crystals of alkali metal intercalated iron selenide of nominal compositions $A_{0.8}(FeSe_{0.98})_2$ (A=K, Rb, Cs) were grown from the melt using the Bridgman method as described in Ref. [1, 16]. Superconducting transition temperature of $Cs_{0.8}Fe_{1.6}Se_2$ was found to be 29.6(2) K by measuring both the electrical resistivity and magnetization [17].

The diffraction data have been taken at ambient temperature with PILATUS 6M detector [18] at ID29 ESRF beamline at wavelength of 0.6887 A; 3600 frames were collected in a transmission geometry and shutterless mode taking 0.1 degree angular spacing between the images; therefore a full sphere up to ca. 1.3 Å of reciprocal space have been explored. Samples were cleaved prior to the measurement and protected from the ambient air by the borosilicate glass capillary with nominal wall thickness of 10 μm; thickness in *c* direction did not exceed 50 μm and lateral size was within 100-150 μm (to be compared to the absorption length of ~80 μm). Comparable data have also been collected for $Rb_{0.8}Fe_{1.6}Se_2$ and $K_{0.8}Fe_{1.6}Se_2$ showing quantitatively similar scattering. The experimental geometry has been refined with the help of CrysAlis software [19] that has also been used for preliminary data evaluation. 3D mapping of scattered intensity and reconstruction of selected reciprocal space layers has been performed with locally developed software. The reconstructed volume is averaged with its symmetrically equivalent orientations employing the Laue symmetry of the average structure, thus improving the signal-to-noise ratio and essentially removing the gaps between individual detector elements.

The IXS experiment was performed on beamline ID28 at the European Synchrotron Radiation Facility. The instrument was operated at 17794 eV, providing an overall energy resolution of 3.0 meV full-width-half-maximum (FWHM). Direction and size of the momentum transfer were selected by an appropriate choice of the scattering angle and the crystal orientation in the horizontal scattering plane. The momentum resolution was set to ~0.25 nm$^{-1}$ × 0.75 nm$^{-1}$ in the horizontal and vertical plane, respectively. Further details of the experimental setup and data treatment procedures can be found elsewhere [20]. Constant-Q scans were performed at room temperature in transmission geometry.

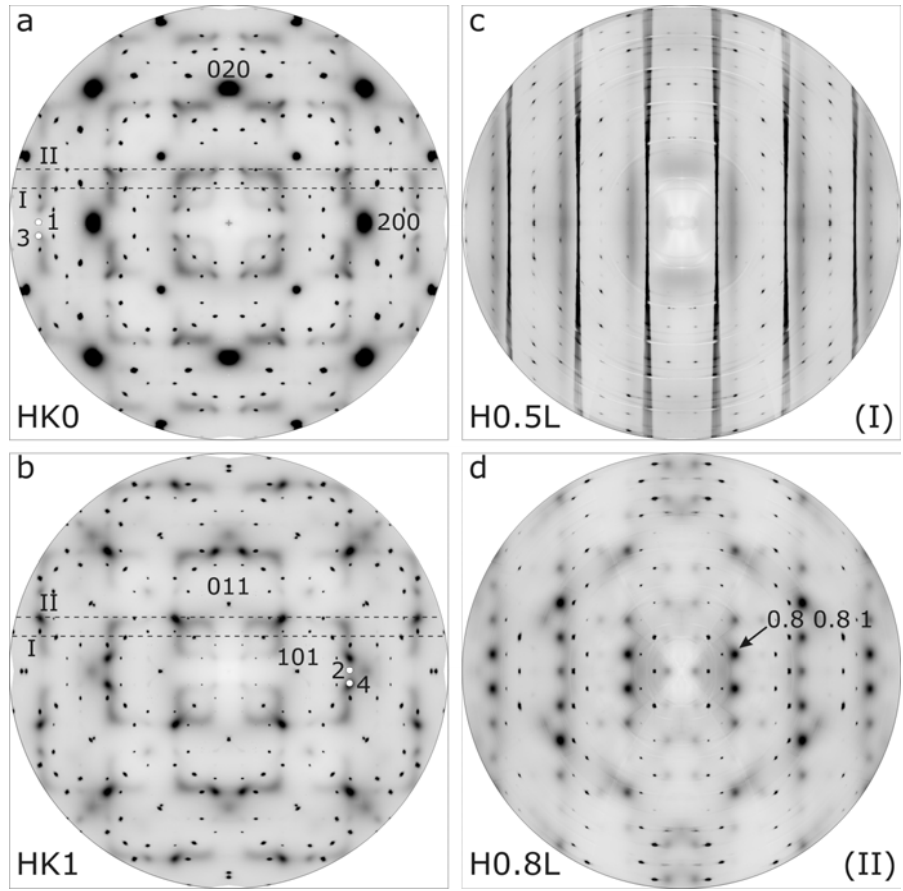

Fig. 1. Reciprocal space cuts for $Cs_{0.8}Fe_{1.6}Se_2$: (a) HK0; (b) HK1; (c) H0.5L; (d) H0.8L. White points denote the Q values for which the IXS experiments were performed; numerical labels are the same as in Fig. 2.

3. RESULTS

Four cuts (layers) of reciprocal space are represented in Fig. 1 with indexation corresponding to the basic average (parent) I4/mmm cell with a = 3.97 Å, c = 15.25 Å. The layers show both sharp Bragg reflections and diffuse intensity, white spots on HK0 and HK1 layers indicate **Q**-vectors where diffuse intensity has further been investigated with inelastic scattering. Inelastic spectra for the given points are shown at Fig. 2 and clearly indicate that for strong diffuse features (labels 3 and 4) an elastic signal largely dominates over the inelastic scattering.

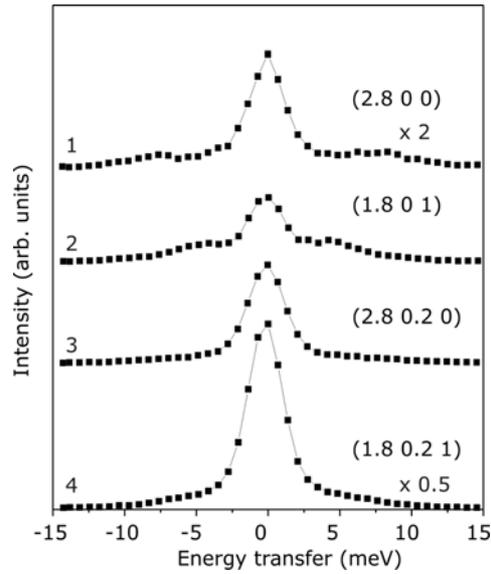

Fig. 2. Selected IXS spectra taken on $Cs_{0.8}Fe_{1.6}Se_2$ crystal. Numerical labels are the same as in Fig. 1; spectra are shifted along the vertical axis for clarity.

Close inspection of distribution of diffracted intensity in reciprocal space allows to identify two phases: one of them (major) corresponds to well-characterized $Cs_{0.8}Fe_{1.6}Se_2$ [2] (I4/m, with $\sqrt{5}\times\sqrt{5}$ superstructure), another one (minor) possesses slightly smaller in-plane parameter as seen from the splitting of the Bragg nodes. The in-plane lattice parameter for this secondary phase is commensurate with the spots normally refereed as a $\sqrt{2}\times\sqrt{2}$ superstructure (Fig. 3 and 4). Note that this superstructure is not commensurate with the main phase.

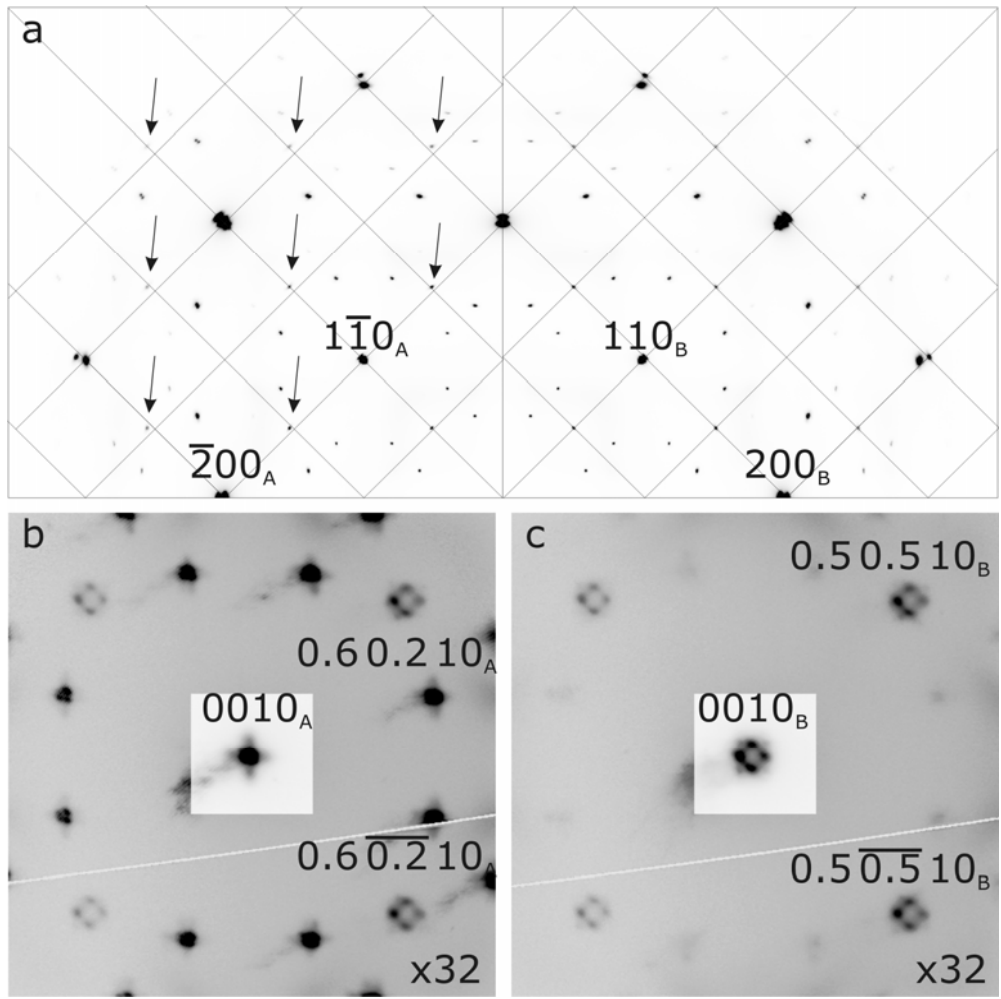

Fig. 3. Reciprocal space cuts parallel to HK0 plane illustrating the splitting of minor phase reflections and Bragg rods. Cut (a) corresponds to HK0 plane; thin lines represent primitive cell for main (left panel) and minor (right panel) phases. Arrows denote diffuse rods incommensurate to the main phase and associated to the minor one. Cuts (b) and (c) cross (0010) reflections of main (denoted as A) and minor (B) phase respectively and show fourfold splitting of both Bragg rods and spots of the minor phase. White arcs correspond to the gaps between the elements of detector. Intensity of Bragg reflections (0 0 10) and their proximity is reduced by factor 32 comparing to the rest of cuts b and c.

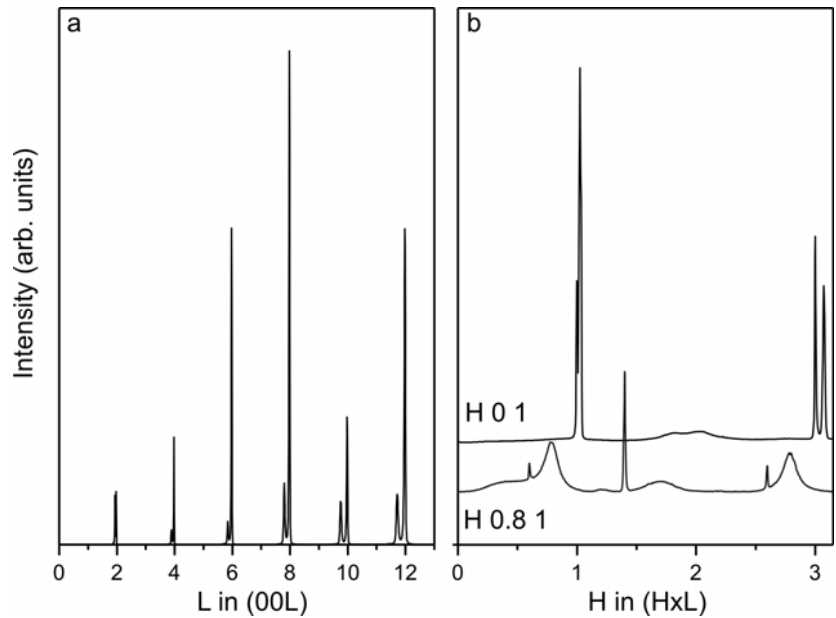

Fig. 4. Intensity distribution along specific directions: (a) along c*; (b) parallel to a*. H and L scales are given in reciprocal lattice units of main phase.

H0.5L cuts of reciprocal space (Fig.1b) clearly show that this "superstructure $\sqrt{2} \times \sqrt{2}$ spots" are in reality modulated Bragg rods. Moreover, fourfold splitting of the rods becomes well visible on HKL sections with large L (Fig 3b and 3c), it agrees with corresponding splitting of the Bragg reflexes of the minor phase (Fig. 3c) and should be attributed to the symmetry lower than orthorhombic. A monoclinic distortion calculated from the four-fold splitting amounts to $90.7^0$. Such splitting is not an effect of mosaicity since Bragg reflections of the main phase do not manifest fourfold pattern (Fig.3b). An inspection of intensity distribution along $c^*$ (Fig. 4a) shows that the minor phase manifests slightly bigger (~2.4%) $c$ lattice parameter than the major phase. The distribution of intensity along $a^*$ (Fig. 4b, H01 line) illustrates the in-plane compression of the lattice of the minor phase also be ~2.4%, indicating the overall approximate volume compression by 2.4% as compared to the main phase.

Therefore, Bragg diffraction can be decomposed on two patterns. Two-phase mixture has a dominant contribution of the vacancy-ordered phase [2], and an admixture of a second low-symmetry phase. Rod-like scattering observed for the minor phase is indicative of a planar disorder and assumes a layered structure with short range interlayer correlations at best. Assuming similar chemical composition and scattering power for 00L reflections, the sample contains about 10% of the minor phase.

Besides of the Bragg diffraction, an additional diffuse component can be identified (Fig. 1). It appears to be commensurate to the main phase (Fig. 1 and Fig. 4b, H0.81 line). Notably, this diffuse scattering does not show any signatures of the interplane disorder observed for the minor phase (Fig.1b). Comparison of scattering patterns collected under similar conditions in the series $Cs_{0.8}Fe_{1.6}Se_2$ (strong diffuse component) – $Rb_{0.8}Fe_{1.6}Se_2$ (weak diffuse component with the same structure) - $K_{0.8}Fe_{1.6}Se_2$ (hardly visible diffuse component) allows us to attribute the diffuse scattering to a correlated disorder presumably in the layers filled with alkaline ions.

The diffuse scattering data in the proximity of 000 Bragg node were further averaged by translation symmetry of the parent lattice. Symmetrised pattern and the way it fills reciprocal space are shown in Fig. 5a,b. Fourier filtering has been applied to $I(q)^{1/2}$, following power spectra are shown in Fig. 3c,d.

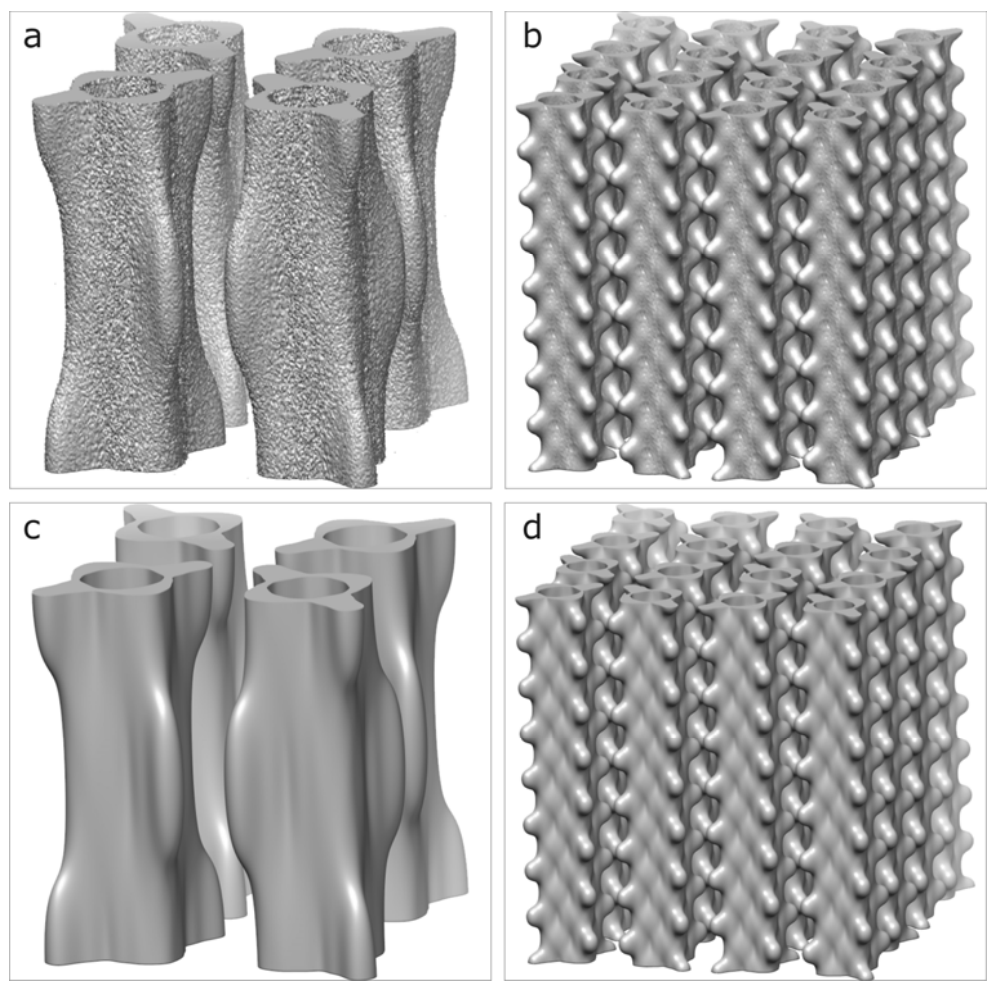

Fig. 5. Basic pattern of diffuse scattering in $Cs_{0.8}Fe_{1.6}Se_2$: (a) $2a^* \times 2a^* \times 2c^*$ volume with stretched c* axis; (b) $4a^* \times 4a^* \times 16c^*$ volume; (c) and (d) Fourier-filtered signal (see text). Origin is situated in the center of cubic volume.

Therefore, taking into account the elastic nature of the diffuse scattering (Fig. 2), it can be attributed to the short range correlations intrinsic to the main phase, presumably linked to off-stoichiometric Cs sublattice.

DISCUSSION

We first like to discuss the appearance of a secondary phase that seems to be an important issue for such type of superconductors, as superconductivity by the major phase is questioned and a phase separation is conjectured. As can be seen from the widths of Bragg reflections (Fig. 4a, 4b), the size of regions of coherent scattering for both phases is of the same order of magnitude. The interlayer correlations in the minor phase are not completely lost, as follows from the slight but apparent modulation along the diffuse rods, coinciding with the periodicity of lattice. In our case the secondary phase most probably relates to planar fragments; lamella-like inclusions coherently intergrown with the main phase may serve as a good candidate. The highest symmetry compatible with the observed diffraction pattern for the minor phase is monoclinic with the monoclinic angle about $90.7^0$. We have seen similar distortion for many samples tested including those with Rb.

Notably, the minor phase has smaller volume comparing to the main one. We have to note that temperature-dependent studies cannot be considered as an ultimate proof of an intrinsic phase separation mechanism as the mobility of alkaline metal is too high and it can be more or less freely redistributed between two phases, thus assuming slightly different chemical compositions. The same argument holds for any surface-sensitive studies, where volatility of alkali metal also becomes crucial, in particular for HV and UHV conditions. A fortiori the surface layer is depleted, and its structural and – obviously – electronic properties will differ from the bulk state.

The diffuse component of scattering is commensurate with the major phase and therefore reflects its intrinsic short-range correlations. This weak scattering component can be expected from non-stoichiometric composition for the $A_xFe_{2-y}Se_2$ family, but has not been reported before. We have to admit that, at variance with Bragg scattering, the interpretation of diffuse scattering is much less straightforward and there is no common recipe on how to treat the data. We therefore start from the interpretation proposed by Krivoglaz [21], with the diffuse

intensity being approximated in terms of Fourier components of concentration fluctuations, $c_\mathbf{q}$:

$$I_{DIF}(\mathbf{q}) \propto \left\langle |c_\mathbf{q}|^2 \right\rangle \left[ (f_\mathbf{q}^A - f_\mathbf{q}^B) - \overline{f}\mathbf{q}A_\mathbf{q} \right]^2 \qquad (1)$$

Here form-factors $f_\mathbf{q}^A$ and $f_\mathbf{q}^B$ stay for two components mixed in the average structure; for the considered case they can be approximated by Cs X-ray scattering factor and zero for a Cs vacancy. The second term in square brackets accounts for a displacive component due to the structural deformation near a defect, for more details see Refs. [11]. The intensity modulation of the observed diffuse scattering with **q** is partially affected by local atomic displacement but the applied translational averaging smears out this dependence, in particular in the proximity of the (0 0 0) node. Therefore, in first approximation the displacive component can be neglected and we can parameterize the intensity as

$$I_{DIF}(\mathbf{q}) \propto \left\langle |c_\mathbf{q}|^2 \right\rangle f_\mathbf{q}^{Cs2} \approx A \left\langle |c_\mathbf{q}|^2 \right\rangle. \qquad (2)$$

The short-range-ordering (SRO) parameters are the Fourier coefficients of $\left\langle |c_\mathbf{q}|^2 \right\rangle$ and can be evaluated from the Fourier transform (further denoted as FT) of the diffracted intensity. Such a Fourier transform represents an analogue of Patterson (autocorrelation) function but for a deviation from the average structure [22].

Instead of standard FT[I(q)], in Fig. 6 we represent of FT[I(q)$^{1/2}$]. Such kind of representation allows reducing the number of descriptors as compared to Warren-Cowley parameters. FT[I(q)] is the self-convolution of FT[I(q)$^{1/2}$], so its significant part is more extended in real space. Practically, only the nearest alkaline metal layer has to be taken into account; much more coefficients have to be retained to reproduce in-plane correlations. On a qualitative level, the correlations for the nearest neighbours in 100 direction are negative, and in 110 they are positive. The reconstruction of the real space approximants for the given system is beyond of the scope of this paper, instead we will speculate on the underlying mechanisms giving rise to that particular shape of diffuse scattering.

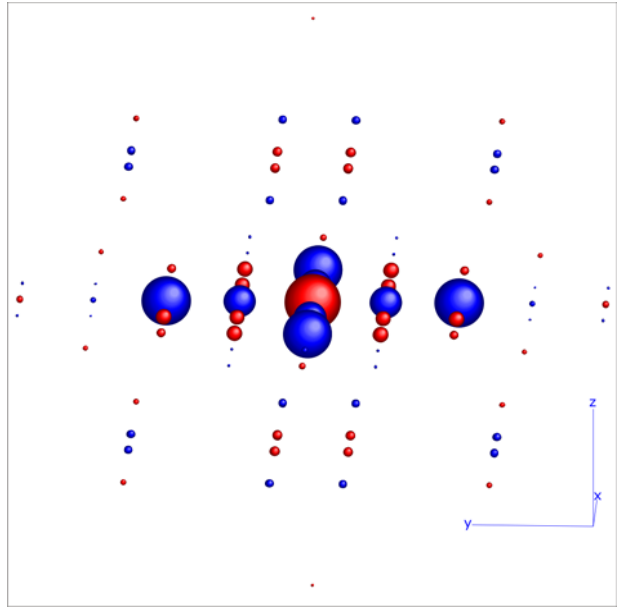

Fig. 6. (Colour online) Fourier transform of $I(q)^{1/2}$ (red (light grey) spheres correspond to positive values, blue (dark grey) spheres – to negative, sphere radius is proportional to the value; zero node is scaled arbitrarily.

As has first been suggested by Krivoglaz [21] and Moss [23], and subsequently demonstrated for the Cu-Pd and Cu-Al alloy systems [24], the fine structure of diffuse scattering due to the short range ordering (SRO) is sensitive to details of the electronic structure. In order to see to what extend such a link could be seen in the title compound, we attempt to correlate observed diffuse scattering with the shape of the Fermi surface. The physical background for such correlation could be tentatively attributed to a perturbation of inter-atomic interactions at a certain set of **q**-vectors corresponding to effective nesting of the Fermi surface; such perturbation would manifest itself in the spectrum of compositional fluctuations, mapped in turn by the diffuse scattering.

The direct calculation of the Fermi surface in $A_xFe_{2-y}Se_2$ compounds is far from being trivial due to the intrinsic non-stoichiometry, eventual disorder, large true unit cell (as opposed to small average parent structure) and the necessity to implement spin-polarized calculations. In the usual approximations, the Fermi surface is considered as containing two nested corrugated tubes propagating along c* and centred around $(\pi,\pi,q_z)$ line, together with pockets centred around point $(0,0,2\pi)$ [25]. As seen from Fig. 1 and Fig. 6a, the diffuse scattering shows a set features that may deal with underlying nesting, namely: i) diffuse pattern consists of corrugated tubes, centred around $(\pi,\pi,q_z)$ line; ii) the system of tubes shows I-centring; iii) disorder-related component of diffuse scattering is practically absent in H0L and 0KL planes

(and their translation equivalents). These features cannot be reproduced by an intraband nesting of corrugated tubes or an interband nesting of two tubes, as in these cases the nesting construction would appear as set of tubes centred around Γ-Z lines, that is $(0,0,q_z)$. Diffuse scattering in turn can be decomposed to the set of anisotropic Lorentzian peaks (Fig. 6c) and tube-like pattern without sharp features (Fig. 6d); at least the second component can be related to the interband nesting of tubes with small pockets, lying on the $q_z$ axis, presumably at $(0,0,2\pi)$. For the evaluation purpose we performed non-polarized electronic structure calculations with the Perdew-Burke-Ernzerhof generalized gradient approximation (PBE-GGA [26]) using a Wien2k package [27]. For calculations, Brillouin zone of the average I4/mmm structure was divided in a k-mesh of the 32x32x16 size. All atomic positions were set as fully occupied. As an estimate of the nesting construction we used $|\text{grad}\chi_\mathbf{q}|$, where $\chi_\mathbf{q} = \sum_\mathbf{k} [n_F(\varepsilon_\mathbf{k}) - n_F(\varepsilon_\mathbf{k+q})]/(\varepsilon_\mathbf{k} - \varepsilon_\mathbf{k+q})$ is the real part of the bare spin susceptibility (Lindhard function) at $\omega \to 0$; $n_F(\varepsilon) = 1/[\exp(\varepsilon/k_BT)+1]$ denotes the Fermi-Dirac distribution function.

Starting tube-like fragments of Fermi surface are shown in Fig. 6e together with nesting construction taking the shape of corrugated tubes (Fig. 6f) and their sections (Fig. 6b), to be compared to diffuse component shown in Fig. 6d. The extent of pockets should be limited in $(q_x,q_y)$ plane in order to moderate blurring of the nesting construction; it should be also limited in $q_z$ direction, otherwise *I*-centering of the resulting pattern would be suppressed and disappear completely if a pocket transforms to a thin tube. In fact, even so-called "hidden" nesting [28], without a formation of real pocket, may be sufficient.

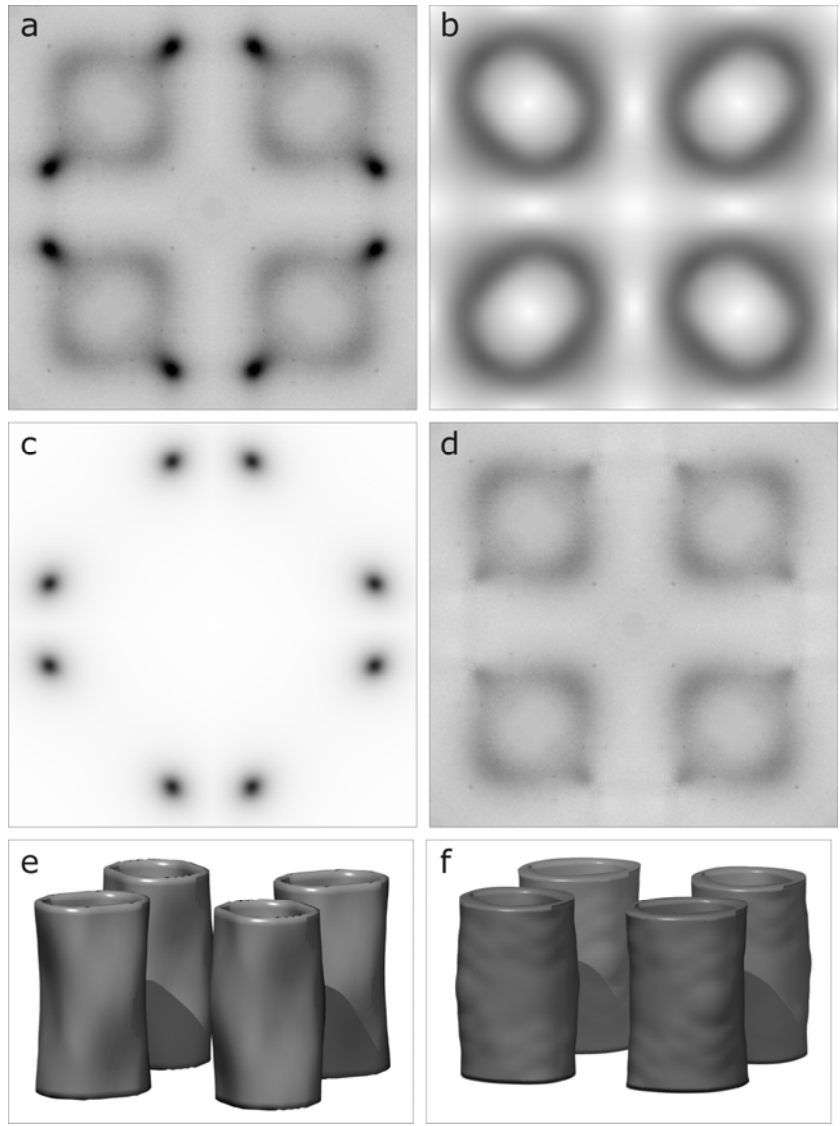

Fig. 6. Symmetrised pattern of diffuse scattering in HK0 plane: (a) periodic 2a*×2a* unit, decomposed to the Lorentzian peaks at (0.1 0.1 1) positions (b) and continuous pattern (d), in comparison to the nesting construction (see text). Selected sheet of Fermi surface (e) is shown together with the nesting construction (f).

CONCLUSION

We show, with the help of diffraction mapping of reciprocal space that inspected $Cs_{0.8}Fe_{1.6}Se_2$ single crystals in fact consist of two coexisting intergrown phases. One of them is well described by the modulated I4/m phase with partial ordering of Fe vacancies [2], while the second (minor) phase has lower symmetry (not higher than monoclinic) structure with weakly correlated planar fragments. Our experiments have also revealed a structured diffuse scattering mostly related to a disorder in Cs sublattice as follows from observed weakening of the X-ray diffuse signal in the series A=Cs, Rb, K. The diffuse scattering forms 3D objects in reciprocal space that can be, at first approximation, parameterized with Cs compositional fluctuations. Correlation pattern indicate relatively strong negative correlations in [100] direction and weaker positive correlations in [110] direction; correlations between disordered Cs layers are the smallest in this hierarchy and do not propagate beyond the nearest layer. Elastic nature of the observed scattering evidenced by inelastic X-ray scattering experiment excludes any effect of potentially soft phonons. We suggest that observed diffuse signal contrasts a perturbation of inter-atomic interactions at a certain set of q-vectors corresponding to effective nesting of the Fermi surface. Detailed analysis of the observed diffuse features together with other probes of Fermi surface and *ab initio* calculations should help to reveal more details about the electronic structure of novel $A_xFe_{2-y}Se_2$ superconductors.

We conclude with a note on a bulk character of all our observations. At variance with the surface-sensitive probes, the diffraction from the sample with the characteristic size of the order of the absorption length represents its intrinsic structural characteristics and allows one to minimize the artifacts related to the surface properties and/or sample preparation issues. We intentionally omit the discussion whether major or minor phases are superconducting or not, as this problem cannot be solved solely by means of diffraction. Instead we provide a constraining input for further discussion, whether it will be based on the experimental measurements or *ab initio* calculations.

**Acknowledgments**

A.K.M. acknowledges the support by the Scientific Exchange Programme Sciex-NMSch (Project Code 10.048) and E.P. acknowledges the support by the NCCR MaNEP Project.